# Universal Entanglement and an Information-Complete Quantum Theory

*Zeng-Bing Chen*

National Laboratory of Solid State Microstructures and School of Physics, Nanjing University, Nanjing 210093, China
E-mail: zbchen@nju.edu.cn

Keywords: universal entanglement, information-completeness, quantum gravity, quantum information, quantum trinity

The most challenging problem of modern physics is how to reconcile quantum theory and general relativity, namely, to find a consistent quantum theory in which gravity is quantized. This Perspective focuses on such a tentative theory called the information-complete quantum theory (ICQT), in which (1) spacetime (gravity) as a physical quantum system plays a central role for formulating the theory, and (2) there are no any classical systems and concepts. Here universal spacetime-matter entanglement “glues” spacetime and matter (matter fermions and their gauge fields) as an indivisible trinity, encodes information-complete physical predictions of the world, and is as universal as universal gravitation. After summarizing the basic theoretic structure of the ICQT, conceptual advances achieved so far and some new issues within the ICQT are considered. While such a theory integrating quantum gravity is of fundamental interest to a wider audience, its relevance to quantum information technologies is discussed, with emphasis on its potential impacts on quantum computing and quantum communication.

## 1. Introduction

Modern paradigm of physics is firmly built upon quantum theory and general relativity, which impact current society broadly and deeply. As general relativity is a classical theory, it is in embarrassment that we describe the same world with the two totally different theories, each being extremely successful in its scale of validity. Thanks to the tradition of unifying originally distinct physical theories or phenomena, unification of quantum theory and general relativity is perhaps the first-love problem of many, if not all, theorists. However, the historical developments show us that the problem is extremely difficult. Up to now, we have two main competing theories, superstring theory[1] and loop quantum gravity (LQG),[2][3][4][5] to target the problem. The two theories represent of course the heroic efforts in our time for achieving ultimately a unified understanding of nature. Meanwhile, their own theoretical progresses, while having remarkable results, are still on the way. It is a kind of comfort that we have the Standard Model, the tremendous successes of modern quantum field theory. Unfortunately, the Standard Model, founded decades ago, does not describe gravity; the experimental search of new physics beyond the Standard Model[6] is depressing—The definite exception is the mystery of dark energy[7] and dark matter.[8]

Thus, we are now in an unfortunate situation in the sense that there is no direct empirical information or data to motivate a deeper theory of nature. The only way to proceed is to seek clues from the logic and conceptual contradictions among empirically successful theories. Below let us list three of such clues (for a more comprehensive list, see Smolin's book[9]), which are relevant to this Perspective.

*Dark energy and dark matter*—One of the biggest cosmological mysteries is whether or not dark matter and dark energy exist. Their existence is, however, the most economic and natural way of understanding the astronomical observations.

*Quantizing general relativity*—If we hold a firm belief that our world is genuinely quantum, we have to find a quantum theory of general relativity, whose essential theoretic structure, e.g., the background-independence,[1][3][4][5] must be kept. In this regard, LQG is most attractive and will be intensively used in this Perspective.

*"Classical remnants" in quantum theory*—A logically consistent resolution of quantum measurement problem[10][11][12][13][14][15] still remains a challenge. The evolution of quantum states is broken into two pieces corresponding to *two* postulates of quantum mechanics: a unitary evolution for the probability amplitude (Schrödinger's equation) and a non-unitary state collapse associated with the probability (quantum measurement).

This mysterious breakdown of unitarity requires a macroscopic, thus classical, "apparatus". Therefore, current quantum mechanics "contains classical mechanics as a limiting case, yet at the same time it requires this limiting case for its own formulation".[16] Namely, current quantum theory has classical remnants, such as classical measuring apparatuses and observers, beyond the description of its formulation. It should be emphasized that nothing is changed regarding the measurement postulate and related probability description of current quantum formalism even in the superstring theory and LQG. If the *M*-theory,[1] the updated theory of superstring, is designed to be the "Theory of Everything" including the whole Universe, what is the external observer who collects various probabilities of events occurring within the Universe and what is the external observer's apparatus measuring the Universe? Without changing the theoretic structure of current quantum theory, even the *M*-theory fails to assign a logically meaningful wave function to the whole Universe.

According to the above-mentioned clues, one of the logic possibilities is to reconsider the very foundation of quantum mechanics and ask: Could the logic and conceptual contradictions among our current theories be rooted in the incompleteness of quantum theory?

Anyone who dares to challenge the completeness of current quantum mechanics should keep in mind the following facts. First of all, it is Einstein, together with Podolsky and Rosen (EPR), who first questioned the completeness of quantum mechanics, well-known as the EPR paradox.[17] The follow-up discover of Bell's inequalities[18][19] and their experimental tests[20][21][22][23][24] give us an impression that quantum mechanics wins[24] against the EPR argument (local realism). Second, quantum mechanics is so successful a theory and so robust a theoretic structure; even a very tiny nonlinear extension[25] of Schrödinger's equation immediately leads to inconsistencies.[26][27][28] As will be shown soon, it is a giant conceptual step, but not a tiny one, that might make sense for a deeper theory of nature.

Here we provide an argument against the information-completeness, rather than the completeness, of current quantum theory. Then we give an overview of the information-complete quantum theory[29][30] (ICQT) suggested recently as an extension of current quantum formalism. For the ICQT to be a consistent theoretic structure, spacetime (gravity) has to be treated as a physical quantum system. Spacetime-matter entanglement "glues" spacetime and matter (matter fermions and their gauge fields) as an indivisible entity and is argued to be as universal as universal gravitation. Conceptual advances achieved so far and some new issues within the ICQT are considered. The relevance of the ICQT to quantum information technologies[20][31][32] is discussed, with emphasis on the potential impacts on quantum information science.

## 2. Information-incompleteness of current quantum theory

As any information carriers are physical systems, physics is certainly about information. In quantum realm, it is then necessary to require that *any information must be carried and acquired by certain quantum systems that encode their complete information*. Based on this basic requirement (called *the information-completeness principle* hereafter) that any information-complete quantum description of nature has to satisfy, we argue below that current quantum theory is information-incomplete.

"Unperformed experiments have no results".[33] Thus quantum states of a single quantum system provide no information, which can only be accessed by certain quantum interaction. Let us consider a $d_{\mathcal{S}}$-dimensional quantum system $\mathcal{S}$ in a state

$$|s,\mathcal{S}\rangle = \sum_{j=1}^{d_{\mathcal{S}}} a_j |j,\mathcal{S}\rangle \tag{1}$$

where $\sum_{j=1}^{d_{\mathcal{S}}} |a_j|^2 = 1$ and $|j,\mathcal{S}\rangle$ is an eigenstate, associated with an eigenvalue $s_j$, of an observable $\hat{s}$ of system $\mathcal{S}$. Another $d_{\mathcal{A}}$-dimensional ($d_{\mathcal{A}} \geq d_{\mathcal{S}}$) system $\mathcal{A}$ (called a "quantum apparatus") in an initial state $|0,\mathcal{A}\rangle$ is used to interact and thus measure $\mathcal{S}$. For an ideal measurement, known as pre-measurement proposed by von Neumann,[34][35] the interaction between $\mathcal{S}$ and $\mathcal{A}$ induces a unitary transformation such that

$$\hat{U}_{\mathcal{S}\mathcal{A}}(\hat{s},\hat{r})|0,\mathcal{A}\rangle|s,\mathcal{S}\rangle = \sum_{j=1}^{d_{\mathcal{S}}} a_j|j,\mathcal{S}\rangle|j,\mathcal{A}\rangle \tag{2}$$

which is a standard form of the two-party entangled pure state. Here $|j,\mathcal{A}\rangle$ is the eigenstate of $\mathcal{A}$'s observable $\hat{r}$ with the eigenvalue $r_j$. For latter convenience, we call $\hat{U}_{\mathcal{S}\mathcal{A}}(\hat{s},\hat{r})$ the measurement operation and $(\hat{s},\hat{r})$ the observable pair regarding this specific measurement. For definiteness, we suppose that the entangled state in Equation (2) has been Schmidt-decomposed ($a_j > 0$). Now imagine a simple universe where there are only two systems $\mathcal{A}$ and $\mathcal{S}$. For them to acquire information the only way is simply to interact and entangle with each other. The resulting entanglement encodes all the acquired information—Each state of the Schmidt bases $\{|j,\mathcal{S}\rangle\}$ and $\{|j,\mathcal{A}\rangle\}$ records an output of the observable pair $(\hat{s},\hat{r})$, while $|a_j|^2$ is the corresponding probability if information contained in the entangled state is quantified (actually, uniquely quantified) by the usual entanglement entropy for pure states.[36][37][38] These outputs and probabilities are physical predictions. If there is no entanglement, there is no information and thus no physical predictions.

Now we can make an important observation based on the above consideration: In a genuinely quantum world, *entanglement is necessary and sufficient for acquiring information*, i.e., physical predictions. In this sense, entanglement is *the* measurement in the quantum realm. If one has, instead, a set of observable pairs $(\hat{s}_p,\hat{r}_p)$ with $p = 1,2,3,...$, the entangled states created during these measurements read

$$\hat{U}_{\mathcal{S}\mathcal{A}}(\hat{s}_p,\hat{r}_p)|0,\mathcal{A}\rangle|s_p,\mathcal{S}\rangle \equiv |p,\mathcal{S}\mathcal{A}\rangle \tag{3}$$

where $|s_p,\mathcal{S}\rangle = \sum_{j_p=1}^{d_{\mathcal{S}}} a_{j_p}|j_p,\mathcal{S}\rangle$. Of course, mutually complementary properties cannot be measured accurately at the same time, a fact known as the complementarity principle.

In current quantum theory, the picture is, however, different; pre-measurement is not an orthodox quantum measurement. There, one needs an apparatus with a large number of degrees of freedom to interact with system $S$ as in various quantum measurement models.[11][12][13][14] Whatever the apparatus-system interaction could be, such an orthodox quantum measurement must contain certain classical elements—The Bohr's Copenhagen interpretation simply describes the measuring apparatus and observer classically.[16][39] Otherwise, there is no way of switching from a unitary process to a non-unitary state collapse. Nevertheless, the measurement postulate (Born's probability description) is absolutely wanted to link the microscopic quantum system and the macroscopic outputs, namely, to make physical predictions. The price to pay is the mysterious breakdown of unitarity because of the state collapse. In other words, the measurement postulate is a necessary bridge between the coherently evolving quantum world and the classical predictions, i.e., the classical data of the outputs and the associated probabilities about certain observable.

In conventional quantum measurement, one can measure various observables by choosing bases for these observables, as shown in Equation (3). The choice of the basis information is totally observer's free will[40] or freedom of choice.[41][42] The basis information is classical and neither carried by any quantum system, nor described by the theory. Thus, via the measurement postulate we are asking classical questions to the quantum world. But, why do we ask classical questions when the world is quantum in essence?

On one hand, general relativity teaches us that *spacetime is a dynamical and physical system*, and the same thing as gravity.[3] This immediately implies that there are, even in principle, no perfectly isolated systems[43] as they must live in and couple with dynamical spacetime. Without quantizing spacetime, traditional quantum theory must be incomplete. On the other hand, LQG provides a mathematically rigorous framework of quantizing spacetime. Yet, the theoretic structure of current quantum theory keeps unchanged there; the interpretational and conceptual problems of quantum foundations still puzzle us.

To sum up, current quantum theory does not satisfy the information-completeness principle stated above. It is an "unfinished revolution" [3] as there is certain information not carried by any physical quantum system, or there are classical terms/concepts not well justified by the quantum formulation. Even in LQG (as well as in superstring theory, of course), the interpretational and conceptual problems of quantum foundations remain there—They root deeper than quantization of spacetime and call for a major conceptual step for their resolution. In other words, *current quantum theory is itself an unfinished revolution, not simply because spacetime is not quantized, but rather because its own formulation calls for a radical change*.

## 3. Information-complete quantum theory

If current quantum formulation is information-incomplete, what is the information-complete formulation? Below we will show that within the ICQT, it is possible to keep the unitary quantum evolution as a universal law, while making physical predictions without any classical terms or concepts such as probability and external observers. What enables this is the information-completeness principle. Einstein's two major discoveries, the concept of entanglement (discovered together with Podolsky and Rosen) and general relativity, play a core role within our theory.

### 3.1. An information-complete trinity

Let us continue to consider the simple universe consisted of two systems $\mathcal{A}$ and $\mathcal{S}$. Without invoking observers or having to appeal to any concept like free will, what is the quantum system that encodes the basis information if measuring $\mathcal{S}$ along various bases? Obviously, one has to introduce the third $d_{\mathcal{P}}$-dimensional quantum system $\mathcal{P}$ (called the "programming system") that interacts with both $\mathcal{A}$ and $\mathcal{S}$ and encodes the basis information. The unitary transformation generated by the interaction is denoted by

$$\hat{U}_{\mathcal{P}(\mathcal{SA})} = \sum_{p=1}^{d_{\mathcal{P}}} |p,\mathcal{P}\rangle\langle p,\mathcal{P}|\, \hat{U}_{\mathcal{SA}}(\hat{s}_p,\hat{r}_p) \tag{4}$$

Now if system $\mathcal{P}$ is prepared in an initial state $|0,\mathcal{P}\rangle = \sum_{p=1}^{d_{\mathcal{P}}} g_p |p,\mathcal{P}\rangle$, the whole system ($\mathcal{P}$, $\mathcal{A}$, and $\mathcal{S}$) will be entangled from the initial state $|0,\mathcal{P}\rangle|0,\mathcal{A}\rangle|s_p,\mathcal{S}\rangle$ via the $\hat{U}_{\mathcal{P}(\mathcal{SA})}$ operation as follows

$$|\mathcal{P},\mathcal{SA}\rangle \equiv \sum_{p=1}^{d_{\mathcal{P}}} g_p |p,\mathcal{P}\rangle\, |p,\mathcal{SA}\rangle \tag{5}$$

which is a two-party ($\mathcal{P}$ vs. $\mathcal{SA}$) entanglement. Here we always assume that $|\mathcal{P},\mathcal{SA}\rangle$ has been Schmidt-decomposed. In this case $\{|p,\mathcal{P}\rangle\}$ is the "programmed basis".

We call the above structure the *information-complete trinary description*, which needs some comments as following.

1) Though we call $\mathcal{S}$ the system and $\mathcal{A}$ the apparatus, their roles in entanglement $|p,\mathcal{SA}\rangle$ are completely the same—They are mutually measuring because of entanglement. The physical predictions about them are encoded by the Schmidt form of $|p,\mathcal{SA}\rangle$, implying that $d_{\mathcal{A}} = d_{\mathcal{S}} \equiv d$ for an ideal information-complete description; higher dimensions of any single system ($\mathcal{S}$ or $\mathcal{A}$) are not accessible by entanglement.
2) For a $d$-dimensional quantum system, its information-complete set of operators[44][45][46] contains $d^2$ elements/observables. Thus for the programming operation $\hat{U}_{\mathcal{P}(\mathcal{SA})}$ to be information-complete, we must have $d_{\mathcal{P}} \geq d^2$ such that the above-mentioned basis information can be completely encoded by $\mathcal{P}$. Similarly to the above comment, the symmetric role of $\mathcal{P}$ and $\mathcal{SA}$ implies $d_{\mathcal{P}} = d^2$, as well as a mutually measuring relation between $\mathcal{P}$ and $\mathcal{SA}$.
3) $|\mathcal{P},\mathcal{SA}\rangle$ contains entanglement (*dual entanglement*) at two distinct levels: the $\mathcal{P}$-$\mathcal{SA}$ entanglement between $\mathcal{P}$ and $\mathcal{SA}$, and the "programmed entanglement" $|p,\mathcal{SA}\rangle$ between $\mathcal{S}$ and $\mathcal{A}$, as programed by $|p,\mathcal{P}\rangle$. Dual entanglement thus encodes information-complete physical predictions of the whole system (the "quantum trinity")—It is necessary and sufficient for acquiring *complete* information (Recall the previous statement that entanglement between $\mathcal{S}$ and $\mathcal{A}$ is necessary and sufficient for them to acquire information, but not complete information).

4) The role of $\mathcal{P}$ is different from that of either $\mathcal{S}$ or $\mathcal{A}$. In particular, observables defined in the Hilbert spaces of $\mathcal{P}$ and $\mathcal{S}\mathcal{A}$ are information-complete, while those defined in the Hilbert space of either $\mathcal{S}$ or $\mathcal{A}$ are information-incomplete.

From the above discussion, the information-complete trinary description arises as an unavoidable consequence of the information-completeness principle stated above. For this reason, sometimes we can call such a description itself as the information-completeness principle.

### 3.2. Gravity enters

The ICQT excludes single, isolated physical systems in our conventional sense from the very beginning, as they are simply meaningless for acquiring information. As argued above, a trinity is necessary from the information-completeness principle. But why is it also sufficient? In conventional quantum measurement model, there is the so-called von Neumann chain, which means that one is always allowed to introduce more and more other quantum systems interacting with the trinity; the chain is unlimited in principle. Amazingly, if the programming system $\mathcal{P}$ is spacetime (i.e., gravity) being a quantized physical system, the von Neumann chain is terminated and the sufficiency of the our trinary picture immediately follows, as *there is no spacetime beyond spacetime*. Now something very beautiful happens here: *Spacetime/gravity must be quantized not merely in its own right, but rather is required to be a programming quantum system by the ICQT for its own consistent formulation*. Indeed, we have the most remarkable trinity of nature—matter fermions ($\mathcal{S}$), their gauge fields ($\mathcal{A}$), and gravity (spacetime); the Higgs field could be related to the extra components of an elementary gauge field in an information-complete unified theory[47] in (9+1)-dimensional spacetime. *Implementing the* $\widehat{U}_{\mathcal{P}(\mathcal{SA})}$ *operation demands the* $\mathcal{P}$-$\mathcal{SA}$ *coupling, which is provided in nature exactly by the gravity-matter coupling*.

For latter convenience we introduce some notations here. Matter fermions are described by Dirac's field $\hat{\psi}(x)$ as usual. They also form the irreducible representation of a Yang-Mills gauge group $G$, whose generators are $\{T_L, L, M, \ldots = 1,2,\ldots,\dim(G)\}$ and related gauge fields $\hat{C}_\mu(x) = \hat{C}_\mu^L(x)T_L$. A spacetime coordinate $x = (x_\mu)$, where $\mu, \nu, \ldots = 0,1,2,3$ are spacetime tangent indices. The gravity is described by the spin connection $\widehat{\omega}_{\mu J}^I$ and the tetrad field $\hat{e}_\mu^I(x)$; the metric tensor $\hat{g}_{\mu\nu}(x) = \eta_{IJ}\hat{e}_\mu^I(x)\hat{e}_\nu^J(x)$. Indices $I, J, \ldots$ label the Minkowski vectors and the Minkowski metric $\eta_{IJ}$ has signature $[-,+,+,+]$.

In the Hamiltonian formulation of LQG,[1][3][4][5] the 4-dimensional spacetime manifold $\mathcal{M}$ is foliated into a family of spacelike hypersurfaces $\Sigma_t$, labelled by a time coordinate $t$ and with spatial coordinates on each slice: $g_{\mu\nu}x^\mu x^\nu = -N^2(x^0)^2 + g_{ab}(dx^a + N^a dx^0)(dx^b + N^b dx^0)$, where $N$ ($N^a$) is the lapse function (the shift vector) and $a, b, \ldots = 1,2,3$ are the spatial indices. Let $i, j, k, \ldots = 1,2,3$ take values in the *su*(2) Lie algebra (with three generators $\tau_i$). The dynamical variables, known as the Ashtekar-Barbero connection $A_a^j$ and its canonically conjugate momentum $E_j^a$, are defined by[48][49][50]

$$A_a^j = \Gamma_a^j + \gamma K_a^j, \qquad E_j^a = \sqrt{\det g}\, e_j^a \tag{6}$$

Here the 3-metric reads $g_{ab} = \delta_{ij}e_a^i e_b^j$ in terms of cotriad $e_a^i$ (with inverse $e_i^a$); the *su*(2) spin connection $\Gamma_a^j$ is defined by $\partial_a e_b^j - \Gamma_{ab}^c e_c^j + \epsilon_{jkl}\Gamma_a^k e_b^l$ with $e_b^j$ and the Christoffel connection $\Gamma_{ab}^c$; $K_a^j$ is the extrinsic curvature of $\Sigma_t$; $\gamma > 0$ is known as the Immirzi parameter.[50] The canonical conjugate pair $(A_i^a, E_b^j)$ satisfies the standard Poisson bracket $\{A_i^a(x), E_b^j(x')\} = 8\pi G\gamma\delta_b^a\delta_i^j\delta(x - x')$, with $G$ being the Newton constant. Upon quantization, the canonical pair $(A_i^a, E_b^j)$ is promoted into operators $(\hat{A}_i^a, \hat{E}_b^j)$. To simplify our notations, we omit the Higgs field and the conjugate variables of $\hat{C}, \hat{\psi}$, and only write down the total action of the trinary fields formally

$$S(\hat{A}, \hat{E}; \hat{C}, \hat{\psi}) = S_G(\hat{A}, \hat{E}) + S_{G\text{-}M}(\hat{A}, \hat{E}; \hat{C}, \hat{\psi}) \tag{7}$$

whose explicit forms are irrelevant here and can be found in many reviews and books.[1][3][4][5][51]

### 3.3. Postulates

After the above preparations, we summarize the basic postulates of the ICQT as what follows.

*The trinity postulate*—Under the information-complete trinary description, the quantum trinity (dividing the Universe into $\mathcal{P}$, $\mathcal{A}$, and $\mathcal{S}$, here the trinary fields for matter fermions, their Yang-Mills gauge fields, and the gravitational field) is necessary and sufficient for providing its complete physical predictions.

*The state-dynamics postulate*—The Universe in $|A; (C, \psi)\rangle$ is self-created from no spacetime and no matter with the least action, namely,

$$|A; (C, \psi)\rangle = e^{iS(\hat{A}, \hat{E}; \hat{C}, \hat{\psi})/\hbar} |\emptyset\rangle$$

$$\delta S(\hat{A}, \hat{E}; \hat{C}, \hat{\psi}) |A; (C, \psi\rangle = 0 \tag{8}$$

where $|\emptyset\rangle \equiv |\emptyset\rangle_{\mathrm{G}} \otimes |\emptyset\rangle_{\mathrm{M}}$ is the common empty state of geometry and matter. (*Remarks*: The requirement of the least action gives the constraint conditions and the equations of motion, all acting upon $|A; (C, \psi)\rangle$.)

*The observable postulate*—All observables and their physical predications are completely encoded by the dual entanglement structure of the trinary fields.

To get an ease understanding of the ICQT, below we illustrate the formulation of our theory by assuming that the programming basis is spanned by the spin-network states in the LQG.[1][3][4][5] We will give physical argument to support the assumption. While the ICQT does not rely on this specific choice of basis, the spin-network programing basis does make the physical picture more transparent.

In LQG, the basic mathematical device is the holonomy $h_e(A)$, which is a path-ordered exponential of the integral over the Ashtekar-Barbero connection along the edge (or, link) $e$, namely, $h_e(A) = P\exp\left[\int_e A_a^i(x)\tau_i dx^a\right]$. The kinematical Hilbert space is constructed by using functions of the holonomies. Then implementing the three constraint conditions of the theory yields finally the physical Hilbert space. A remarkable result of LQG is to identify the state space of quantized gravity, which is spanned by the spin-network states $|\Gamma, (j_1, \ldots, j_L), (i_1, \ldots, i_N)\rangle \equiv |\Gamma, \{j_l\}, \{i_n\}\rangle \equiv |\Gamma, \vec{j}, \vec{\imath}\rangle$. These states are characterized by the triplet $(\Gamma, \vec{j}, \vec{\imath})$ for an abstract graph $\Gamma$ with $N$ nodes $(i_1, \ldots, i_N)$ and $L$ oriented links $(j_1, \ldots, j_L)$ in three-dimensional region $\mathcal{R}$ embedding two-dimensional surface $\mathcal{F}$; a spin-network state $|\Gamma, j_l, i_n\rangle$ is an irreducible $j_l$ representation of $su(2)$ for the link $l$ and the $su(2)$ intertwiner for the node $n$. The spin-network states have two important properties.[3][5] First, they form a complete orthogonal basis,[52] namely, $\langle\Gamma', \vec{j}', \vec{\imath}'|\Gamma, \vec{j}, \vec{\imath}\rangle = \delta_{\Gamma\Gamma'}\delta_{\vec{j}\vec{j}'}\delta_{\vec{\imath}\vec{\imath}'}$. Second, they diagonalize both the area $\widehat{\boldsymbol{A}}(\mathcal{F})$ operator defined for $\mathcal{F}$ and the volume operator $\widehat{\boldsymbol{V}}(\mathcal{R})$ defined for $\mathcal{R}$: $\widehat{\boldsymbol{A}}(\mathcal{F})|\Gamma, j_l, i_n\rangle = \boldsymbol{A}(j_l)|\Gamma, j_l, i_n\rangle$ if $l \in \mathcal{F} \cap \Gamma$ and $\widehat{\boldsymbol{V}}(\mathcal{R})|\Gamma, j_l, i_n\rangle = \boldsymbol{V}(i_n)|\Gamma, j_l, i_n\rangle$ if $n \in \mathcal{R}$.

In the presence of matter[3][5][51] (ignoring again the Higgs field), the holonomy is likewise defined by $h_e(A, C) = P\exp\left[\int_e [A_a^i(x)\tau_i + C_a^L(x)T_L]dx^a\right]$, a spin-network state turns out to be $|\Gamma, j_l, i_n\rangle \otimes |k_l, F_n, w_n\rangle$. Here $F_n$ and $w_n$ are number of fermions (as well as anti-fermions which we ignore here to simplify notations) and the field strength at node $n$, respectively; $k_l$ is flux of the electric gauge fields across surface $l$. We can then expand $|A; (C, \psi)\rangle$ in terms of $|\Gamma, j_l, i_n\rangle \otimes |k_l, F_n, w_n\rangle$, i.e.,

$$|A; (C, \psi)\rangle = \sum_{\substack{l \in \mathcal{F} \cap \Gamma \\ n \in \mathcal{R}}} S_\Gamma(l, n)\, |\Gamma, j_l, i_n\rangle \otimes |k_l, F_n, w_n\rangle = \widehat{U}_{GM}(t)|\emptyset\rangle \tag{9}$$

which is generated via $\widehat{U}_{GM}$ (with an explicit time-dependence) from $|\emptyset\rangle$ and has the same form as Equation (5). This naturally motivates us to assume that $\{|\Gamma, j_l, i_n\rangle\}$ forms the programing basis required in the ICQT. In this case, $\widehat{U}_{GM}$ has a factorizable structure [see Equation (4)]

$$\widehat{U}_{GM} = \sum_{\substack{l\in\mathcal{F}\cap\Gamma \\ n\in\mathcal{R}}} |\Gamma, j_l, i_n\rangle\langle\Gamma, j_l, i_n| \widehat{U}_G(t) \otimes \widehat{U}_{G|M}^{(\Gamma,\vec{j},\vec{\imath})}(k_l, F_n, w_n, t) \tag{10}$$

such that $\widehat{U}_G(t)$ for gravity and the programed unitary operations for matter generate the states as following

$$\widehat{U}_G(t)|\emptyset\rangle_{\mathrm{G}} = \sum_{\substack{l\in\mathcal{F}\cap\Gamma \\ n\in\mathcal{R}}} S_\Gamma(l,n)|\Gamma, j_l, i_n\rangle$$

$$\widehat{U}_{G|M}^{(\Gamma,\vec{j},\vec{\imath})}(k_l, F_n, w_n, t)\ |\emptyset\rangle_{\mathrm{M}} = |k_l, F_n, w_n\rangle \tag{11}$$

The state $|k_l, F_n, w_n\rangle$ also defines a graph (the “matter graph”) with node $n$ and link $l$. Then spacetime-matter entanglement in Equation (9) is actually quantum entanglement between the spacetime graphs and the matter graphs. The relation, if any, between the matter graphs and the Feynman graphs is certainly an interesting problem.

Here we write, again formally, the total Hamiltonians $\widehat{H}_{GM}$ for gravity and matter, $\widehat{H}_G$ for gravity only, and $\widehat{H}_{G\text{-}M}$ for coupling between gravity and matter (Dirac’s field and the Yang-Mills fields) as

$$\widehat{H}_{GM}\big(\hat{A}, \hat{E}; \hat{C}, \hat{\psi}\big) = \widehat{H}_G\big(\hat{A}, \hat{E}\big) + \widehat{H}_{G\text{-}M}\big(\hat{A}, \hat{E}; \hat{C}, \hat{\psi}\big)$$

$$\widehat{H}_{G\text{-}M}\big(\hat{A}, \hat{E}; \hat{C}, \hat{\psi}\big) = \widehat{H}_{Dirac}\big(\hat{A}, \hat{E}; \hat{C}, \hat{\psi}\big) + \widehat{H}_{YM}\big(\hat{A}, \hat{E}; \hat{C}\big) \tag{12}$$

In terms of the spin-network programming basis, our information-complete trinary description leads to

$$\widehat{H}_{G\text{-}M} = \sum_{\substack{l\in\mathcal{F}\cap\Gamma \\ n\in\mathcal{R}}} |\Gamma, j_l, i_n\rangle\langle\Gamma, j_l, i_n| \otimes \widehat{H}_{G|M}^{(\Gamma,\vec{j},\vec{\imath})}(k_l, F_n, w_n) \tag{13}$$

Now it is easy to verify that

$$
\begin{aligned}
&\widehat{H}_{GM}|A;(C,\psi)\rangle = 0 \quad \Rightarrow \quad \frac{\partial}{\partial t}\widehat{U}_{GM} = 0 \\
&i\hbar\frac{\partial}{\partial t}\widehat{U}_{G} = \widehat{H}_{G}\widehat{U}_{G}, \qquad i\hbar\frac{\partial}{\partial t}\widehat{U}_{G|M}^{(\Gamma,\vec{j},\vec{i})} = \widehat{H}_{G|M}^{(\Gamma,\vec{j},\vec{i})}\widehat{U}_{G|M}^{(\Gamma,\vec{j},\vec{i})}
\end{aligned} \tag{14}
$$

$\widehat{H}_{GM}|A;(C,\psi)\rangle = 0$ is the Hamiltonian constraint of the trinary fields, i.e., the Wheeler-DeWitt equation, which shows the disappearance of the Schrödinger-type dynamics (the "problem of time"[3][5][53]) in quantum gravity/cosmology. Here dynamics is timeless for the trinity as a whole so that Schrödinger's equation losses its meaning as describing a dynamical evolution. But due to the dual entanglement structure, our formalism recovers the Schrödinger-type dynamics separately for gravity and for matter and thus provides a solution to the problem of time, similarly to the Page-Wootters mechanism[54][55] and as shown in the second line of Equation (14).

### 3.4. More on the state-dynamics postulate

Kinematics and dynamics in current physics are separated and as such, one needs the initial/boundary conditions plus dynamical laws to make predictions. In sharp contrast, kinematics about states and observables for an individual system is either meaningless for acquiring information or information-incomplete in the ICQT. Seen from the ICQT, quantum states in the usual formulation of quantum theory contain redundant information. By integrating the trinary picture into the theoretic structure and assuming that dual entanglement of the trinity encodes the information-complete physical predictions—No entanglement implies no physical predictions, *the ICQT unifies kinematics and dynamics*, and eliminates all unphysical degrees of freedom and redundant information inherent in current quantum description. Therefore, the unification of kinematics and dynamics arises here as a new feature of the ICQT, as a natural consequence of the information-completeness principle. This is the reason why the second postulate is called the state-dynamics postulate.

## 4. Conceptual applications

In modern physics, no single person changed our world view forever more than Einstein, as he kept most attention to the fundamental issues of physics. On one hand, because of the experimental confirmation of the Bell-inequality violations for many times, the win of quantum theory and the failure of Einstein's attitude to quantum theory were announced from time to time. But we should keep in mind a fact: The Bell-inequality violations are enabled by entanglement, a concept discovered by Einstein, together with Podolsky and Rosen. Only by taking into account the fundamental features (background independence and diffeomorphism invariance) of Einstein's general relativity, can we successfully formulate LQG. As we argued above, Einstein's discovery that spacetime itself is a physical system plays a unique role for consistent formulation of the ICQT. So, if we are humble a little bit, we should never underestimate what Einstein has given us, especially on his style of understanding nature from the first principles and fundamental aspects. In this Section, let us concentrate on the conceptual aspects of the ICQT.

### 4.1. Universal entanglement

Largely due to the rapid advances in quantum information science, quantum entanglement[38] is nowadays even known in the general public. It is usually understood as an exotic and rare quantum phenomenon that is technologically uneasy to observe. It is also a fragile "resource" for quantum information processing.[20][31][32] Let us argue here that entanglement is far more than that—*Entanglement is the building block of the world and universal in exactly the same sense that gravitation is universal*.

Traditionally, we usually study a physical system (ideally, a single particle) by well isolating it from the rest of the Universe. So current physics (classical or quantum) gives certain reality to free systems, single particles, and so on. We all know that this is an approximation, but an extremely good approximation. On the other hand, when we are talking about the physics of spacetime, why this is a safe approximation *always*? Note that in principle, there are no isolated physical systems (the Universe as a whole is the only exception) simply because spacetime itself is a physical system coupling universally with any form of matter, or, gravitational force "can never be shielded."[43] Since such a universal coupling establishes quantum entanglement between spacetime and matter, entanglement must be universal, too. *Universal spacetime-matter entanglement is the information-complete physical predictions of the ICQT and glues spacetime and matter*. Thus, the concept of universal entanglement lies not only at the heart of the ICQT, but also at the heart of the world.

Remarkably, the $\mathcal{P}$-$\mathcal{SA}$ entanglement between $\mathcal{P}$ and $\mathcal{SA}$ is a two-party entanglement, although having three constituents ($\mathcal{P}$, $\mathcal{S}$, and $\mathcal{A}$). As indicated by the ICQT, such a two-party entanglement is enough for understanding nature. The deep reason why this is the case might relate to the comprehensibility of nature without the need of multiparty entanglement at the most fundamental level—Multiparty entangled states generally do not allow a Schmidt-decomposition and have much richer structure, currently not fully understood,[38] than two-party ones.

### 4.2. A new world view

The world view underlying the ICQT is certainly different from our current understanding in various aspects, some of which will be discussed below.

#### *4.2.1. Quantum determinism*

In current quantum theory, while the dynamical evolution governed by Schrödinger's equation is deterministic and unitary, a quantum state in a coherent superposition is non-unitarily collapsed upon measurement into an eigenstate of the measured observable; the collapse is genuinely random and only probability about its occurrence can be predicted. On the contrary, dynamics of the trinity in the ICQT is deterministic and unitary and results in dual entanglement encoding all observables and their information-complete physical predictions. Such quantum determinism encoding an information-complete reality of the world is to be compared with classical determinism, which means the deterministic change of a single possibility of a physical system. For quantum determinism, all allowed possibilities are coherently stored in the entangled structure and deterministically changed.

*4.2.2. Quantum relationalism*

Because of the basic property of the Schmidt decomposition, pure-state two-party entanglement is invariant under local unitary transformations. This is true for both the $\mathcal{P}$-$\mathcal{SA}$ entanglement and the programmed entanglement. The different choices of local bases for any one of the trinary systems only correspond to the different observables of that system. However, the relations between the observables for the relevant two parties keep invariant. To compare with the relational interpretation[3][56] of quantum mechanics, we would like to say that, in a genuinely quantum world, entanglement is a relation, and actually *the* relation—It is not the constituent parts of the Universe, but rather their relations/entanglement that are/is of physical significance.

*4.2.3. Local versus global eyes*

According to the formulation of the ICQT, the Universe can be viewed from both local and global eyes. On one hand, the Universe state $|A;(C,\psi)\rangle$ (see below for a particular example) determined by dynamics contains the information-complete physical predictions of the whole Universe and is of course a global description. On the other hand, in solving the problem of time, we can recover the Schrödinger-type dynamics for both spacetime/gravity and matter. Particularly, a matter graph (a "*spacetime-matter atom*" as it is programmed by spacetime) satisfies the standard Schrödinger equation with a matter Hamiltonian $\widehat{H}_{G|M}^{(\Gamma,\vec{j},\vec{\imath})}(k_l,F_n,w_n)$.

Thus, both eyes, one with time and another without time, can detect the same physics of the Universe. This picture, as implied by the ICQT, seems to be precisely stated by Weyl[57]: "The objective world simply is, it does not *happen*. Only to the gaze of my consciousness, crawling along the lifeline of my body, does a section of this world come to life as a fleeting image in space which continuously changes in time."

*4.2.4. Quantum world for information*

The tradition of doing quantum physics is to start from a classical theory and then to quantize it. However, if we believe that the world is truly quantum, there is no classical theory; we have classical physics simply because we got physical laws first from macroscopic scales, which seem to be "classical". The ICQT does not rely in any way on the concepts of probability and observer, or any related classical concepts in its own formulation; a probability description and related classical notions appear only as an information-incomplete and approximate description of nature. While classical physics is of course a useful clue for seeking ultimately quantum theory, quantization of classical Hamiltonian or action often suffers from the ambiguity (e.g., operator ordering) problem,[2][3] in which various forms of a Hamiltonian operator result in the same classical theory. Here an interesting problem arises: For a physicist who knows nothing about classical physics but has a quantum theory like the ICQT, how can he/she get the correct and unique form of quantum Hamiltonian or action?

Meanwhile, the world described by the ICQT is also all about information encoded by a dual entanglement structure. Thus, information is indeed a fundamental notion for understanding nature, as Wheeler's famous thesis "it from bit"[58] implies. Looking nature from Wheeler's information thesis indeed leads to many fruitful insights[59][60][61][62] into the structure of quantum theory. Surely, one should notice the marked difference between the information thesis and the ICQT, which talks about complete information encoded in universal spacetime-matter entanglement.

## 4.3. Self-defining quantum structure

The Universe described by the ICQT is self-defining and self-explaining via its trinary constituents, but not defined and explained via any external observers. The dual entanglement pattern among the trinary systems realizes this theoretic structure: Matter and spacetime/gravity are mutually defining and measuring via the $\mathcal{P}$-$\mathcal{SA}$ entanglement; the programmed entanglement between matter fermions and their gauge fields, as programed by gravity, realizes similarly mutually defining and measuring structure. Therefore, quantum states appearing in the $\mathcal{P}$-$\mathcal{SA}$ entanglement and the programmed entanglement are physical predictions of the theory.

Such a self-defining and self-explaining structure, respected by dynamics, has profound physical consequences. In the context of LQG, the geometry operators $\widehat{\boldsymbol{A}}(\mathcal{F})$ and $\widehat{\boldsymbol{V}}(\mathcal{R})$ are "partial observables".[3][5] If the spin-network states indeed span the programing basis in the ICQT, they will be guaranteed to be physical (and thus solve all the constrains of the theory) due to the mutually measuring structure between geometry quanta and matter excitations. Now the programing basis defines a joint property of spacetime and matter. Thiemann[1] gave a mathematical argument showing that the geometry operators become diffeomorphism invariant and thus physical as soon as they couple with matter excitations. Spacetime-matter entanglement is the underlying physics of Thiemann's argument.

The only exception of the self-explaining structure is the common empty state $|\emptyset\rangle \equiv |\emptyset\rangle_{\mathrm{G}} \otimes |\emptyset\rangle_{\mathrm{M}}$ of geometry and matter, which is the initial condition of the dynamical evolution in the ICQT. Note that the empty-matter state $|\emptyset\rangle_{\mathrm{M}}$ is not the conventional field vacuum state, whose definition relies on a background spacetime. If the vacuum energy is to reproduce the cosmological constant ($\Lambda$) term in Einstein's equation, the predicted $\Lambda$ according to various models will be about 51-117 orders[7] of magnitude larger than the observed $\Lambda$. Meanwhile, the empty-geometry state $|\emptyset\rangle_{\mathrm{G}}$ does not correspond to the usual flat spacetime, which is meaningless in the ICQT simply because there is no matter to define it. This hints the

possibility of mathematics (in particular, geometry) based on relational sets, whose elements can only be defined in the relational contexts—If our world is relational, we need relational mathematics.

**4.4 New restrictions of describing nature**

LQG does not explain the matter content of nature, namely, there is no any restriction imposed by LQG on the matter fields as well as their interactions. Any form of the Standard Model matter can be well filled into the formulation of LQG. This is comprehensible as LQG is a mathematically consistent quantization of spacetime only. Unifying matter fermions and their forces is the goal of other unified theories,[6][63][64][65] such as the Grand Unification Theory, superstring, and supergravity. However, such a tendency of unification results in uncontrollable larger and larger symmetries, more and more exotic new particles, and additional physics.[3] While scale fields were added into physics by hand to target many problems, only the Higgs field seems to be supported by nature as a building block, without knowing the fundamental reason behind its success. Is there any fundamental principle that could restrict our ability to arbitrarily introduce, by hand, new physical degrees of freedom, or new fields, or new symmetries?

Certainly, the information-complete trinary picture puts stronger and more restrictions on physical description of nature than the usual formalism, as can be readily seen from our above discussions. Here let us go some further on this issue. Above we have considered the simple universe consisted of two finite-dimensional quantum systems $\mathcal{A}$ and $\mathcal{S}$. The requirement that physical predications are entanglement of $\mathcal{A}$ and $\mathcal{S}$ restricts their dimensions; more dimensions are inaccessible by entanglement. Now suppose that $\mathcal{A}$ are the Yang-Mills fields and $\mathcal{S}$ the fermion fields, both of finite degrees of freedom. A similar argument then implies that both kinds of matter fields should have the same numbers of degrees of freedom. Otherwise, there will be redundant degrees of freedom that are not accessible via entanglement. Of course, when gravity is considered, these redundant degrees of freedom for matter sector can entangle gravity only and thus be interpreted as dark matter or dark energy. But the more economical way is to consider first the simpler case where there are no redundant degrees of freedom. Let us emphasize once again that the physical predictions of the ICQT is dual entanglement and consequently, any redundant information or degrees of freedom beyond dual entanglement is meaningless and should not be introduced into our description. Following this line of thought, we showed, in a primary study,[47] that the restriction of the information-complete trinary description seems to uniquely result in a unified theory of matter and spacetime, provided that spacetime is $(9+1)$-dimensional. The $(3+1)$-dimensional effective theory is promising to reproduce the main features of the Standard Model, plus the dark-matter candidates.

## 5. Quantum-cosmological issues

Obviously, the ICQT has no conceptual obstacle for its applications to quantum cosmology as it describes a self-explaining universe without external observers observing the Universe. Related to some recent results,[66][67][68][69][70][71] we derived[30] previously the variational holographic relation (the holographic relation[72][73][74] was proposed by ’t Hooft and Susskind as a guiding principle of quantum gravity) from a particular form of spacetime-matter entanglement. Proceeding a step further, a more universal form of the Universe state was conjectured to be

$$|\mathrm{Univ}\rangle = \sum_{\substack{l\in\mathcal{F}\cap\Gamma \\ n\in\mathcal{R}}} \frac{e^{-\boldsymbol{A}(j_l)/2\boldsymbol{A}_0-\boldsymbol{V}(i_n)/2\boldsymbol{V}_0}}{\sqrt{Z_\Gamma}} |\Gamma, j_l, i_n\rangle \otimes |k_l, F_n, w_n\rangle \tag{15}$$

where $\boldsymbol{A}_0$ and $\boldsymbol{V}_0$ are two constants to be determined, and $Z_\Gamma$ a normalization constant. When there is a small perturbation to $|\mathrm{Univ}\rangle$, the reduced density operators for matter and for gravity are $\rho'_M = \rho_M + \delta\rho_M$ and $\rho'_G = \rho_G + \delta\rho_G$, respectively. The change of entanglement entropy caused by the perturbation is $\delta\mathcal{E}_{GM} = -\mathrm{tr}(\delta\rho_G \ln\rho_G) = -\mathrm{tr}(\delta\rho_M \ln\rho_M)$. By noting $\mathrm{tr}(\delta\rho_M) = 0$ and defining $\boldsymbol{\delta A}$=δ$\langle\widehat{\boldsymbol{A}}(\mathcal{F})\rangle$ and $\boldsymbol{\delta V}$=δ$\langle\widehat{\boldsymbol{V}}(\mathcal{R})\rangle$, we have explicitly

$$\delta\mathcal{E}_{GM} = \frac{\boldsymbol{\delta A}}{\boldsymbol{A}_0} + \frac{\delta\boldsymbol{V}}{\boldsymbol{V}_0} \tag{16}$$

On the other hand, as shown previously,[30] $|k_l, F_n, w_n\rangle$ is the eigenstate of $\widehat{H}_{G|M}^{(\Gamma,\vec{j},\vec{\iota})}$, namely, $\widehat{H}_{G|M}^{(\Gamma,\vec{j},\vec{\iota})}|k_l, F_n, w_n\rangle = E_{G|M}^{(\Gamma,j_l,i_n)}|k_l, F_n, w_n\rangle$. We separate $\widehat{H}_{G|M}^{(\Gamma,\vec{j},\vec{\iota})}$ and $E_{G|M}^{(\Gamma,j_l,i_n)}$ into two parts

$$\widehat{H}_{G|M}^{(\Gamma,\vec{j},\vec{\iota})} = \begin{cases} \widehat{H}_{G|M}^{(\Gamma,\vec{j}=\emptyset,\vec{\iota})} \\ \widehat{H}_{G|M}^{(\Gamma,\vec{j}\neq\emptyset,\vec{\iota})} \end{cases} \Rightarrow E_{G|M}^{(\Gamma,j_l,i_n)} = \begin{cases} \boldsymbol{E}_{G|M}^{(\Gamma,j_l=\emptyset,i_n)} \\ \tilde{E}_{G|M}^{(\Gamma,j_l\neq\emptyset,i_n)} \end{cases} \tag{17}$$

According to a quantum-information definition, energy without link excitations ($j_l = \emptyset$), i.e., energy related merely to node/volume excitations, is dark energy, while energy with link excitations ($j_l \neq \emptyset$) is the usual "visible energy" (matter fermions and their gauge fields mutually measuring and visible). The sum of the two kinds of energy is called the "entanglement energy"[30] $\Xi_\Gamma = \boldsymbol{E}_{G|M}^{\Gamma} + \tilde{E}_{G|M}^{\Gamma}$, whose variation is

$$\beta\delta\Xi_\Gamma = \beta(\delta\boldsymbol{E}_{G|M}^{\Gamma} + \delta\tilde{E}_{G|M}^{\Gamma}) = \frac{\boldsymbol{\delta A}}{\boldsymbol{A_0}} + \frac{\delta\boldsymbol{V}}{\boldsymbol{V}_0} \tag{18}$$

Now we assume (1) $\boldsymbol{\delta A}/\boldsymbol{A}_0$ is identical to that given by the variational holographic relation, (2) the Hamiltonian[51] $H_\Lambda = -\frac{\hbar c\Lambda}{8\pi l_P^2}\int_R d^3x\sqrt{\det g}$ ($l_P$: the Planck length) related to the cosmological constant term after quantization is contributed solely by the dark energy defined by the ICQT (Keep in mind that energy related to $H_\Lambda$ and dark energy are both volume/node energy), and (3) $\beta^{-1} = \kappa_B T_U = \frac{\hbar a_E}{2\pi c} \equiv \frac{\hbar}{2\pi t_E}$ ($\kappa_B$: the Boltzmann constant, $T_U$: the Unruh temperature[75][76] related to the acceleration $a_E$ of the Universe). The three conditions immediately lead to

$$\boldsymbol{A}_0 = 4l_P^2, \quad \boldsymbol{V}_0 = \frac{4l_P^2}{t_E c\Lambda} \tag{19}$$

This determines the two free constants in $|\mathrm{Univ}\rangle$.

Let us discuss the physical consequences of the above results as follows.

*Einstein's equation as classical limit*—In a thermodynamic argument,[30] dark energy given by the ICQT corresponds to a universal negative pressure $P_U = -\frac{c^4 \Lambda}{8\pi G}$ expanding the Universe with a constant acceleration $a_E$. Combining Jacobson's thermodynamic argument,[77] the ICQT, *for the first time, reproduces the full information of Einstein's equation*, i.e., the classical Einstein equation with the cosmological constant term. The universality of $|\mathrm{Univ}\rangle$ hints the possibility that $\boldsymbol{V}_0$ or $a_E$ might be a universal constant characterizing our Universe, just like $\Lambda$.

*The Universe is not strictly holographic*—The universal relation between entanglement entropy and geometry in Equation (16) modifies the usual variational holographic relation $\delta\mathcal{E}_{GM} = \boldsymbol{\delta A}/\boldsymbol{A}_0$. The extra term $\boldsymbol{\delta V}/\boldsymbol{V}_0$ implies that the Universe is not strictly holographic due to dark energy. As we previously showed,[30] the existence of dark energy is an unavoidable consequence of the ICQT.

*Time's arrow*—In a quantum-computing interpretation of Equation (15), the Universe, as an information-complete quantum computer, has monotonically increasing entanglement entropy, increasing a universal amount, as given by Equation (16), for each computing step. This defines thus an arrow of time.

As another application to quantum cosmology, let us briefly mention the presence of universal entanglement in models of quantum cosmology. Cosmological spacetime is usually highly symmetric due to spatial homogeneity, thus resulting in only finite number of dynamical variables.[78] As far as there are matter fields, as should be, coupled with the cosmological spacetime, the physical Hamiltonian has the same structure as in Equation (12). Therefore, the presence of universal entanglement keeps as a robust and unavoidable feature of our description. The physical consequences of universal entanglement in the context of quantum cosmology will be considered elsewhere.

## 6. Interfacing quantum gravity and quantum information

Recent years witnessed remarkable mutual impacts between quantum gravity and quantum information. Now, quantum entanglement is a routine language in the context of quantum gravity. Entanglement between the inside and the outside of the Rindler horizon is associated with the black-hole entropy.[67][68][69][70][71] Other examples include the studies of entanglement in the context of spin networks,[52][79][80] and holographic entanglement entropy from the Anti-de Sitter/Conformal Field Theory (AdS/CFT) correspondence.[81][82][83] Quantum error correction[84][85][86] was used recently to reinterpret the AdS/CFT correspondence. Actually, a reasonable name of the ICQT itself is quantum information/entanglement dynamics of spacetime and matter.[30] Because the interface of gravity and information is a fertile testbed of quantum gravity, below we discuss some applications of the ICQT at this interface.

### 6.1. Self-computing quantum Universe

The ICQT allows us to define an information-complete quantum computer[29] with a trinary structure and potential of outperforming conventional quantum computers. To interface the present quantum theory of spacetime and matter with quantum information, let us consider the quantum-computing interpretation[30] of the dynamical evolution of the ICQT. $\widehat{U}_G(t)$, called the "spacetime gates", prepares from $|\emptyset\rangle_{\mathrm{G}}$ the spacetime state $\widehat{U}_G(t)|\emptyset\rangle_{\mathrm{G}}$, which acts as the controlling or programing state for the matter sector. Then the "spacetime-matter gates", namely, the controlled-$\widehat{U}_{G|M}^{(\Gamma,\vec{j},\vec{i})}$ operations $\sum_{n\in\mathcal{R},l\in\mathcal{F}\cap\Gamma}|\Gamma,j_l,i_n\rangle\langle\Gamma,j_l,i_n|\otimes\widehat{U}_{G|M}^{(\Gamma,\vec{j},\vec{i})}(k_l,F_n,w_n)$, together with the spacetime gates given above, generate the overall spacetime-matter entangled state $|A;(C,\psi)\rangle$. The "matter gates" $\widehat{U}_{G|M}^{(\Gamma,\vec{j},\vec{i})}(k_l,F_n,w_n,t)$ completely determine the entangled states of matter fermions and their gauge fields, as shown in Equation (11).

Interestingly, we can use $\Gamma_T$ ($T = 0,1,2, ...$) to label the computing steps, as well as the total energy of spacetime or matter for a given graph $\Gamma_T$. $\Gamma_T$ defines actually a discrete time. The computing proceeds from the empty state corresponding to $T = 0$ and consumes matter of increasing energies step by step, resulting in expanded spacetime and more matter described by $|A; (C, \psi)\rangle_\Gamma$. Entanglement entropy of $|A; (C, \psi)\rangle_\Gamma$ is monotonically increasing with $T$, a kind of time's arrow as discussed above. During the information-complete computing process spacetime and matter mutually "borrow" energies while keeping the total energy of spacetime and matter exactly zero. Thus, *our Universe, if indeed being described by the ICQT, is consistently self-computing from nothing and nowhere*. In other words, the Universe is itself a realization of the information-complete quantum computer defined by the ICQT.

### 6.2. Black holes as quantum information processor

Let us consider the limit on the information content of a spacetime region $\mathcal{R}$ associated with an external surface $\partial\mathcal{R}$ of area $\boldsymbol{A}$. The celebrated holographic principle[72][73][74] gives a limit on the entropy of the physical systems within $\mathcal{R}$ as

$$\text{entropy}_{\mathcal{R}} \leq \frac{\boldsymbol{A}}{4l_P^2} \tag{20}$$

For a Schwarzschild black hole the entropy is maximal and this maximal entropy is known as the Bekenstein-Hawking entropy.[87][88][89] By microstate counting,[1][3][5] LQG could explain the holographic bound given in Equation (20). The ICQT gives a rather different picture of the holographic bound and quantum black hole, as we will show immediately.

Entanglement entropy with respect to dual entanglement of spacetime and matter for the region satisfies an obvious relation

$$\mathcal{E}_{GM} \leq \ln D_G \tag{21}$$

where $D_G$ is dimensions of the available spacetime states in $\mathcal{R}$, i.e., the Schmidt number of the spacetime-matter entangled state associated with $\mathcal{R}$. If we require that for a Schwarzschild black hole, $\mathcal{E}_{GM}^{max} = \ln D_G$, the black hole must be in the *maximally information-complete* state

$$|\Gamma, \mathrm{BH}\rangle = \sum_{\substack{l \in \mathcal{F} \cap \Gamma \\ n \in \mathcal{R}}} \frac{1}{\sqrt{D_G}} |\Gamma, j_l, i_n\rangle \otimes |k_l, F_n, w_n\rangle \quad (22)$$

Here $\mathcal{F}$ is still the two-dimensional surface embedded in $\mathcal{R}$. Maximal information-completeness means maximal dual entanglement: Both spacetime-matter entanglement and the programed matter entanglement are maximal. Also, as we stated above, the node/volume degrees of freedom have to be included in $|\Gamma, \mathrm{BH}\rangle$, as our Universe is not strictly holographic. Maximal entanglement has an intriguing property called monogamy:[90] If two parties are maximally entangled, then they can no longer be entangled with any third party. Monogamy enables a conceptually clear understanding of the black hole: Quantum black hole is "information-black", does not allow extracting its information in any way, and thus represents certain *entanglement death* of spacetime and matter defining it.

In quantum information science, entanglement is usually a resource for various quantum information tasks,[20][31][32] such as quantum computing and quantum teleportation[20][31][32][91] (Teleportation of entanglement is known as entanglement swapping.[92]) Maximal dual entanglement possessed by quantum black hole could well be the resource for quantum information processing with spacetime and matter.[93][94]

Classical general relativity is featured by an unavoidable existence of spacetime-matter singularities in black holes (as well as at the Big Bang). Black-hole physics is thus a unique interface of general relativity and quantum theory, hoping to resolve the singularity problem, together with the information-loss problem.[95][96] For the two problems, here we only make general remarks, leaving detailed considerations in the context of the ICQT to be reported elsewhere.[97] *Maximally information-complete entanglement represents a limit ("quantum information-singularity") into which spacetime and matter might evolve. This is the* *fundamental reason why there is no singularity in the sense of classical theory.* As there is no room for non-unitary evolution, there is no information-loss problem, either, in our information-complete trinary description of nature.

### 6.3. Quantum-gravitational attack of quantum cryptography

Quantum communication ultimately aims at absolute security (via quantum cryptography, or quantum key distribution[98][99]) and faithful transfer (via quantum teleportation) of information, classical or quantum. Before considering quantum communication within the framework of the ICQT, let us first make a general remark. All current quantum communication protocols have to make use of classical concepts on information, e.g., classical communication between communicating parties. It is very interesting to see how to formulate communication in the ICQT that does not assume any classical concepts. Quantum teleportation has been discussed above in the context of the ICQT. Now let us consider how to formulate a quantum key distribution process in the ICQT and a possible quantum-gravitational attack to quantum cryptography.

Plaga[100] firstly considered the gravitational attack to quantum cryptography, to the author's knowledge. But without any concrete model of quantum gravity, such an attack could only be formally discussed. On the contrary, the gravitational attack to quantum cryptography can be precisely formulated within the ICQT. To this end, we consider the four-state (generalization to more states is straightforward) BB84 protocol, which works as follows. Alice prepares randomly single-photon (or effectively single-photon) states $|0\rangle \equiv |z_+\rangle$, $|1\rangle \equiv |z_-\rangle$ (in the $z$-basis) and $|\pm\rangle = \frac{1}{\sqrt{2}}(|0\rangle \pm |1\rangle) \equiv |x_\pm\rangle$ (in the $x$-basis). Afterwards, she sends the photon to Bob, who measures it randomly either in the $z$-basis or $x$-basis. The follow-up step is classical post-processing (including basis reconciliation, error correction, and privacy amplification procedures), which demand authentic classical communication between Alice and Bob.

To formulate the ICQT counterpart of the process, keep in mind that we cannot use any classical communication or classical system. To this end, Alice needs to supply a matter particle, to be entangled with the single photon in the $z$-basis and in the $x$-basis; such an entangling in the two bases is enabled and programed by two orthogonal spacetime states ($|z,\mathcal{P}\rangle$ and $|x,\mathcal{P}\rangle$). After entangling, the single photon is sent to Bob and possessed by Bob upon receipt. For convenience, we denote the matter particle (single photon) as system $\mathcal{S}$ (system $\mathcal{A}$). Effectively, the information-complete state possessed by Alice and Bob is then

$$|\text{Alice-Bob}\rangle = \alpha|z,\mathcal{P}\rangle \otimes |z,\mathcal{S};z,\mathcal{A}\rangle + \beta|x,\mathcal{P}\rangle \otimes |x,\mathcal{S};x,\mathcal{A}\rangle \tag{24}$$

where $|\eta,\mathcal{S};\eta,\mathcal{A}\rangle = \frac{1}{\sqrt{2}}(|\eta_+,\mathcal{S}\rangle|\eta_+,\mathcal{A}\rangle + |\eta_-,\mathcal{S}\rangle|\eta_-,\mathcal{A}\rangle)$, with $\eta = x,z$, is a maximally-entangled state. For simplicity, here we take $\alpha = \beta = 1/\sqrt{2}$. The $\mathcal{P}$-$\mathcal{S}\mathcal{A}$ entanglement between $\mathcal{P}$ and $\mathcal{S}\mathcal{A}$ and the programmed entanglement between $\mathcal{S}$ and $\mathcal{A}$, as shown in Equation (24), encode the complete information of relevance, i.e., the perfect correlations of Alice and Bob's bit values (keys) in both bases. The above procedure provides the simplest description of quantum key distribution within the ICQT.

For an eavesdropper, Eve, to attack the key distribution process, she is supposed to have all resources that are allowed by the ICQT. She could use all of her resources to entangle all photons, each in $|\text{Alice-Bob}\rangle$, sent between Alice and Bob; if she keeps Alice's photons, she needs to send Bob replacing photons. After eavesdropping, the final information-complete state of the whole systems allows us to assess the security of quantum key distribution. Of course, currently we still do not know how to define, in the formalism of the ICQT, states for photons and particles as in conventional quantum field theory. Thus, the ICQT counterpart of quantum key distribution and eavesdropping, as given here, is merely tentative; detailed considerations are beyond the scope of this Perspective.

## 7. Discussion and Outlook

Needless to say, the ICQT is of course a kind of quantum theory. However, it integrates the trinary structure occurring in nature, unifies kinematics and dynamics, and assumes that universal entanglement of spacetime and matter encodes the information-complete physical predictions of the world. In this way, it gives up unphysical degrees of freedom and redundant information inherent in current quantum description. The information-complete principle realized via a trinary picture puts stringent restrictions on physical description of nature than the usual formalism; the restrictions might be stringent enough for yielding a unified theory for the matter sector of the world, as well. The ICQT is proposed for resolving the logic and conceptual contradictions among existing theories. Encouragingly, it is the only quantum theory, known so far, that has a correct classical limit (i.e., Einstein's field equation) and defines dark energy with a definite link to the cosmological constant.

Our planet has been getting along with quantum theory for more than a century. Perhaps now *it is time for us to seriously think about how to understand a genuinely quantum world without any classical remnants*, and even to dream quantum information processing with spacetime and matter. The world described by our new theory is universally entangled, quantum-deterministic, quantum-relational, and explains its own existence. The new theory requires spacetime quantization, which plays a central role in its own formulation. This fact might well explain why the theoretic structure like the ICQT hides so deeply. Even if the ICQT is ultimately not the candidate for a unified theory of describing nature, it could well be a start, we hope.

**Acknowledgements**

I am grateful to Xian-Hui Chen, Dong-Lai Feng, Yao Fu, Chang-Pu Sun, and Hua-Lei Yin, among many others, for enjoyable discussions and conversations on various aspects of the theory reported here.

**Data Availability Statement**

No additional data are available in this Perspective.